# Flavor Changing Neutral Currents and Left-Right Symmetric model

H.M.M.Mansour[1] and N. Bakheet
Physics Department, Faculty of Science, Cairo University Egypt
1. mansourhesham@yahoo.com

**Abstract**

In the present work we consider including the Flavor Changing Neutral Currents (FCNCs) in Left-Right Symmetric model. Within this model a simulation calculation are made of the decay width of $Z'_{LR}$ boson, the production cross section of the top quark and the top quark pair production cross section. The events were simulated according to the extension of Standard Model (SM) and the Left-Right symmetry Model (LRSM) based on the gauge symmetry group $SU(3)_C \times SU(2)_L \times SU(2)_R \times U(1)_{B-L}$. The LRSM model exhibits signatures of new physics Beyond the Standard Model (BSM) at the hadrons colliders.

## 1. Introduction

Does the Standard Model of particles physics need an extension? The standard model of electroweak and strong interactions is going to be severely tested as well as many of its extensions that have been proposed to cure its flaws. The observed pattern of neutrino masses the existence of dark matter and the observed matter-antimatter asymmetry are the most severe evidences where the SM fails to explain. It is widely accepted that the SM ought to be extended but no one knows if the proper way has already been explored in the literature. A joint collaboration is therefore needed between the experimental and the theoretical communities. Left-right-symmetric gauge models suggested previously to describe all interactions of elementary particles within a single unified framework are further shown to have the desirable property that maximal parity violation in low-energy weak processes arises purely as a result of spontaneous breakdown of the local gauge symmetry. In the symmetry group $SU_L(2) \times SU_R(2) \times U_{B-L}(1)$; it should be emphasized that B - L is now a local symmetry and not merely a global symmetry. The left- right symmetric model has been thoroughly discussed in the literature [1-4] where it has been shown that it predicts a neutral current which is at present indistinguishable from that of the standard $SU_L(2) \times U(1)$ gauge theory, with almost the same predictions of the $W^{\pm}$ and $Z_L$ masses. The left-right symmetric model differs from the standard model in providing a conceptually different basis for parity violation in weak interactions in that P violation is not intrinsic to the Lagrangian but is spontaneous in nature. With this difference is associated the requirement that the neutrinos in the left-right symmetric model possess finite masses whereas the standard model takes mass less (left-handed) neutrinos.

We search for Flavor Changing Neutral Current (FCNC) mediated by the extra gauge boson $Z'_{LR}$ at the Large Hadrons Collider (LHC) in Left-Right symmetry model [5-7] extension of the Standard Model of particle physics using data produced from Monte Carlo events generators PHYTHIA8 , MADGRAPH5 (matrix elements generators), CALCHEP , LANHEP , and ROOT data analysis . Also we use FeynRules embedded in Mathematica8 and used its result with MadGraph5 matrix elements generators then we produce events collision files Les Houches Event file and use it with Pythia8 to analysis the data of events. Many extensions of the Standard Model predict the extra neutral massive gauge bosons Z'. Z' can induce Flavor Changing Neutral Currents (FCNCs) in higher order correction. In Left-Right Symmetry model one has an extra U (1) group the $Z'_{LR}$ boson which can have a tree-level Z' − q − q' couplings where q is the up-type quarks  (u, c, t) and q' down-type quarks (d, s, b).In our numerical calculations for FCNC in Left-Right Symmetric model we calculate the total cross sections for the signal and the corresponding Standard Model (SM) background processes and decay width.

Figure (1) shows the decay widths of $Z'_{LR}$ boson for the mass range 400 to 2000 GeV in Left –Right symmetry model.

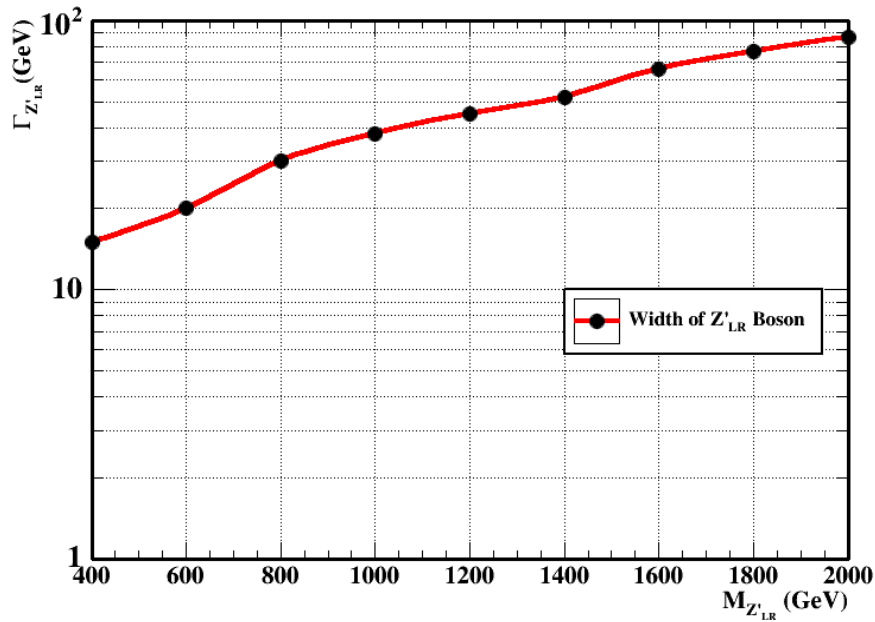

Figure 1: The decay width of $Z'_{LR}$ boson depending on its mass in Left-Right symmetric model at $g_L=g_R=.4$ with the FCNC using MadGraph5 and Pythia8 event generators

We use productions and decay Monte Carlo events simulation of top quark at LHC. The Flavor Changing Neutral Currents (FCNC) couplings of the top quark can be observed.

We use both the single and pair production of top quarks via FCNC interactions through new Left-Right neutral massive gauge boson $Z'_{LR}$ exchange at the LHC ($\sqrt{s}$ = 14 TeV) and we use Left-hand coupling $g_L$ equal to Right-hand coupling $g_R$=0.4 to Study the effects of tree-level FCNC interactions induced by $Z'_{LR}$.

To simulate this work we open the main switches in PYTHIA8:

PartonLevel: all =on

ProcessLevel: resonanceDecays=on allow resonance decays

PartonLevel: ISR=on            allow Initial State Radiation

PartonLevel: FSR=on            allow Finial State Radiation

PartonLevel: MPI=on            allow multiparton interactions

HadronLevel: Hadronize =on     allow Hadronization

*For Left-Right Symmetry model:*

LeftRightSymmmetry: all=off

LeftRightSymmmetry: ffbar2ZR =on

In order to enrich the signal statistics even at the small couplings we consider both $\bar{c}t$ (*top quark anti charm quark*) and $\bar{t}c$ (*anti top quark and charm quark*) productions in the final state.

## 2. The Single Top Quark Production Cross Section
### 2.1. Production Cross Section

Figure (2) shows the cross sections for the process for pp→ ($t\bar{c} + \bar{c}t$) $Z'_{LR}$ LHC for 7teV and 14TeV. Depending on the $Z'_{LR}$ boson mass at $g_L=g_R=.4$

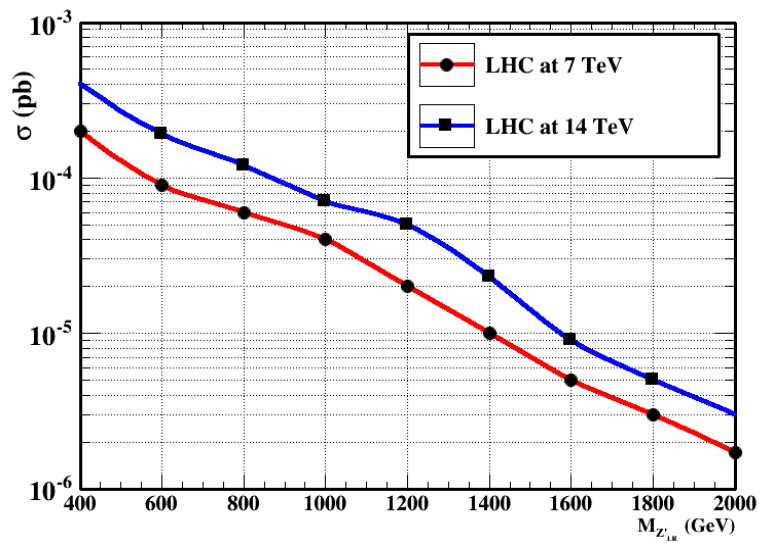

Figure 2: The production cross sections pp→ ($\bar{t}c + \bar{t}c$) $Z'_{LR}$ versus the $Z'_{LR}$ boson mass at the LHC at 7 TeV and 14 TeV at $g_L=g_R=.4$ in Left-Right Symmetric model using MadGraph5 and Pythia8

$Z'_{LR}$ boson contributes through the t-channel and the associated production of single top quarks in the final state $t\bar{c}$ will dominate over the tc final state.
For this process the cross section at √s = 14 TeV is larger than the case at
√s = 7 TeV.

## 2.2 The Rapidity

Figure (3) shows the rapidity distribution of the charm quarks (c and $\bar{c}$) from the signal at the collision energy of 14 TeV.
We sum up the c and $\bar{c}$ distributions where there no clear difference between them.

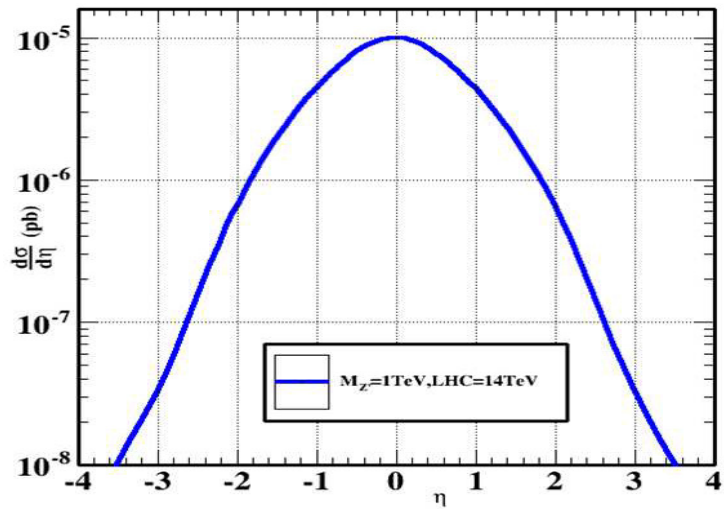

Figure 3: The rapidity distributions of the charm quark at the LHC at 14 TeV at $g_L=g_R=.4$

We apply a rapidity cut | η | < 2.5 for the signal and background analysis. There is a peak in the c-quark rapidity distribution η = 0.The rapidity distribution of final state c quarks in the background
Process pp →($t\bar{c}$ +$\bar{t}c$) $Z'_{LR}$ is shown in figure (4)

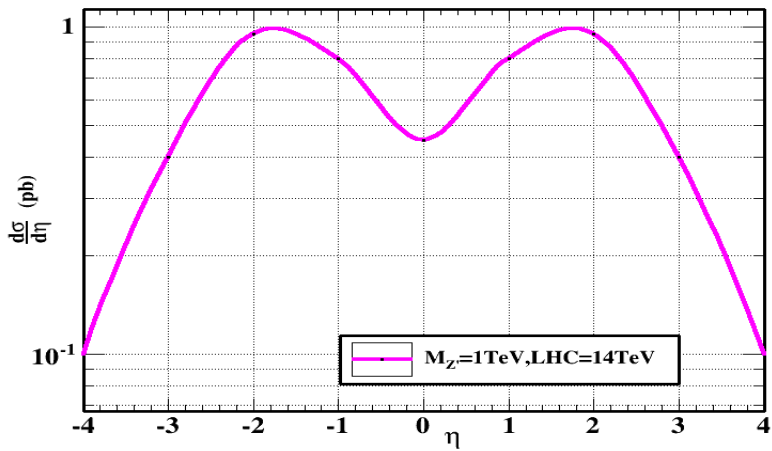

Figure 4: The rapidity distribution of the charm quark for the background process pp →($t\bar{c}$ +$\bar{t}c$) $Z'_{LR}$ at the LHC with 14 TeV

## 2.3 Transverse Momentum

Figure (5) shows the **$P_T$** distributions of the c-quark in the signal process with $M_{Z'}$=1 TeV at the pp center of mass energy of 14 TeV. A high **$P_T$** cut

$$P_T > M_{Z'}/2 - 4\Gamma_{Z'}$$

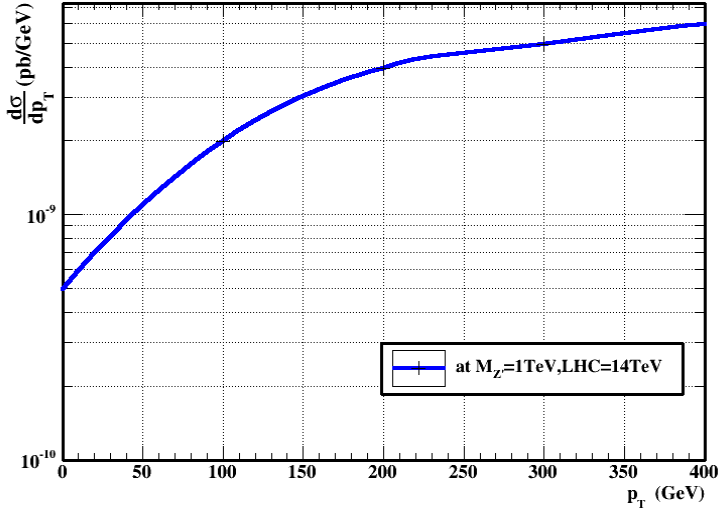

Figure 5: The $P_T$ distribution of the charm quark for the signal process pp $\rightarrow$ ($t\bar{c} + \bar{t}c$) $Z'_{LR}$ at the LHC.

The **$P_T$** distribution of the c-quark from the background is shown in figure (6). We apply these cuts to make analyses with the signal and background.

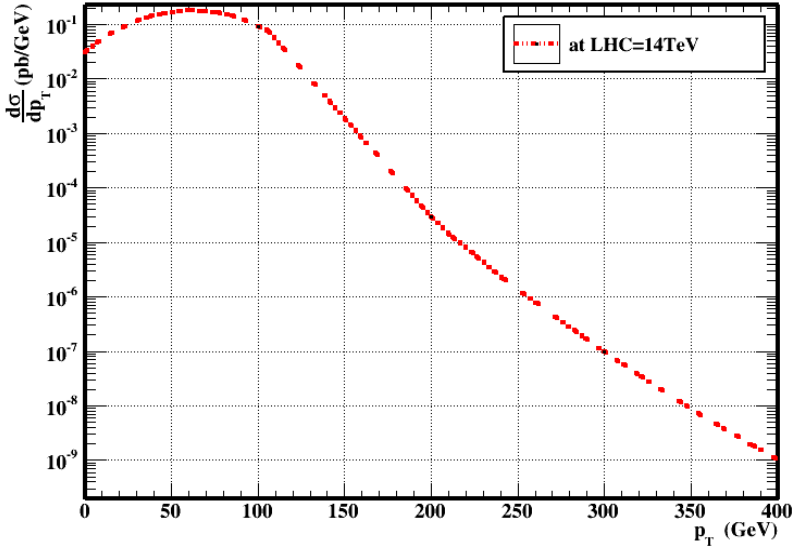

Figure 6: The $P_T$ distribution of the charm quarks for background

## 3. The Top Quark Pair Production

We investigate the top pair production via $Z'_{LR}$ boson exchange at the LHC. The $Z'_{LR}$ boson contributes in the t-channel via the FCNC and in the s-channel through family diagonal neutral current couplings.

## 3.1. Total Production Cross Section

The total cross sections for the top pair production of Z'$_{LR}$ at the LHC are plotted in figure (7). The main background has the same final state as the signal $t\bar{t}$ quarks.

$$pp \rightarrow t\bar{t}Z'_{LR} \rightarrow W^+ W^- b\bar{b} \; Z'_{LR}$$

Where $\quad t \rightarrow W^+ b ; \quad \bar{t} \rightarrow W^- \bar{b}$

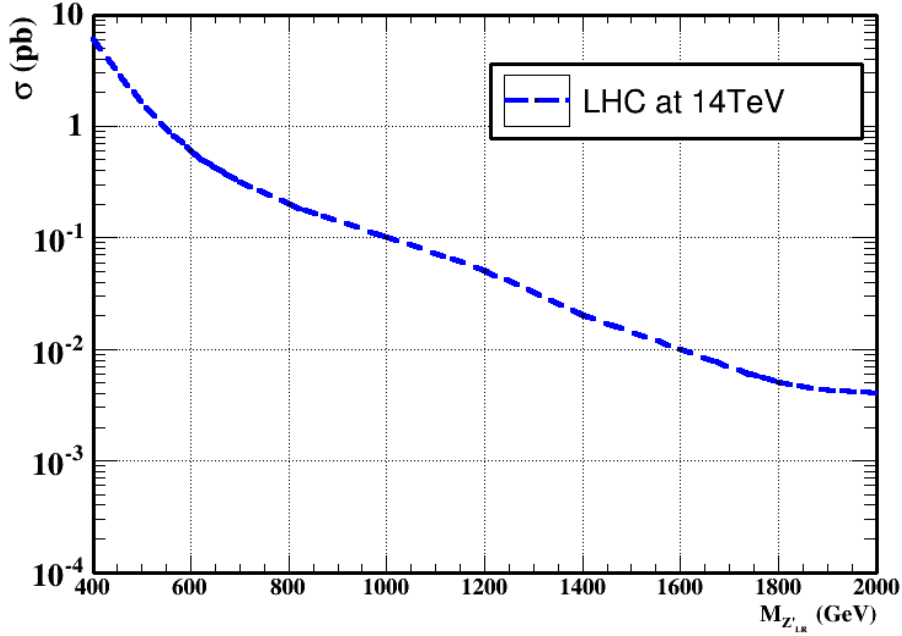

Figure 7: The total cross section for the top pair production at the LHC with 14 TeV depending on the Z'$_{LR}$ mass at $g_L = g_R = .4$

## 4. Conclusion

In the present work we made a simulation calculation of Top Quark Production Cross Section, the rapidity distribution of the charm quarks, the $P_T$ distributions of the c-quark in the signal process. From this work we can say that at tree level in Left right Symmetric Model the single and pair production of top quarks at the LHC can have the Contributions from the couplings of $Z'q\bar{q}$ and the FCNC couplings of $Z'qq'$ (Where q, q' = u, c, t).